# Ordered Arrays of Gold Nanoparticles Crosslinked by Dithioacetate Linkers for Molecular devices

Maryana Asaad,[a]* Andrea. Vezzoli[b], Abdalghani Daaoub[c], Joanna Borowiec[d], Eugenia Pyurbeeva[a], Hatef Sadeghi[c], Sara Sangtarash[c], Simon J. Higgins[b], and Jan. A. Mol[a]

1. School of Physical and Chemical Sciences, Queen Mary University of London, London, UK, E1 4NS, Email: m.asaad@qmul.ac.uk
2. Department of Chemistry, University of Liverpool, Liverpool, UK, L69 7ZF
3. School of Engineering, University of Warwick, Coventry, U.K, CV4 7AL
4. Department of Chemistry, University College London, London, UK, WC1E 6BT

The final performance of a molecular electronic device is determined by the chemical structure of the molecular wires used in its assembly. Molecular place-exchange was used to incorporate di-thioacetate terminated molecules into ordered arrays of dodecanethiol capped gold nanoparticles. X-ray photoelectron spectroscopy confirmed successful molecular replacement. Room-temperature molecular conductance of a statistically large number of devices reveals that conductance is enhanced by up to two orders of magnitude for the di-thioacetate terminated molecules. Density functional theory transport calculations were performed on five different configurations of the di-thioacetate molecules between gold electrodes, and the calculated average conductance values are in good agreement with the experimentally-observed conductance trend. Our findings highlight important cooperative effects of bridging neighbouring gold nanoparticles and choice of appropriate molecular wires when designing devices for efficient transport.

## Introduction

Self-assemblies of molecularly linked gold nanoparticles (AuNPs) have recently attracted significant attention due to their versatile chemical, electrical and optical properties[1-5], making them promising for advanced technologies such as sensors[6,7], optoelectronics[8], biomedical devices[9], thermoelectrics[10] and others.[11,12] The nanoparticles act as nanoscale contact electrodes, and various molecules can be attached to the nanoparticles via anchoring groups.[8] The properties of the network can be tailored by controlling the nanoparticle size and modifying the organic matrix. A wide range of anchor groups have been investigated in self-assembled monolayers (SAMs) studies, but the dominant system in the literature is superlattice structures formed from AuNPs encapsulated with thiols.[13] Despite the availability of significant data on the synthesis, assembly and structure of thiol-functionalized AuNPs, only fundamental concepts have been demonstrated so far. Further improvement of device electrical performance is crucial to for the development of the next generation of nanodevices that have valuable real-world function.

In this work, we explore the incorporation of a new set of dithioacetate (di(SAc))-terminated molecules into SAM AuNPs. The structures of these molecules consist of α-ter- and α-bi-thiophene and phenyl central moieties connected to alkyl chains as illustrated in Figure 1. Single molecule studies have shown that molecular conductance is largely influenced by electronic structure of the molecule.[14-16] We thus chose these three different central units on the basis of their extent of delocalization, going from a less conjugated phenyl-based linker to the well-conjugated α-ter-thiophene moiety. The use of dithioacetate termini allows linkage to two different gold nanoparticles via strong S-Au covalent bonding, which helps improving compactness and robustness of the 2D array and thus the final conductance of the array.[17] The electrical properties of these ligands have been previously explored at the single-molecule level, providing a good reference for comparison to self-assembled monolayers of AuNPs experiments.[18] In this study, the authors found that the presence of the conjugated moiety in the alkyl tunneling barrier facilitates transport through coupling to additional electronic states located on the thiol ending group, which gives a rise to resonance close to Fermi energy. Here, gold nanoparticles are initially stabilized by dodecanethiol (DDT), which is a poor conductor. We then exploit molecular place exchange to insert the di(SAc)-terminated molecules into self-assembled DDT-capped AuNPs. This approach is likely to preserve the size of nanoparticles and does not interfere with the structural order of the network.[8,13,19] X-ray Photoelectron spectroscopy (XPS) confirmed the successful insertion of di(SAc) into the 2D arrays. The room temperature conductance of a statistically large number of devices was measured in air before and after exchange. We observed that upon exchange, the conductance of the network increases by up to two orders of magnitude for the conjugated compounds. Experiment and theory were employed to understand the conductance behavior of the di(SAc) compounds. We find that the single-molecule central unit effects can be translated into SAMs of AuNPs. The presence of conjugated central unit in the molecular backbone influences the S-Au coupling and therefore the final conductance of the array. The magnitude of the conductance increase depends on the nature of the conjugated part. This offers an excellent platform for building nanoscale molecular devices that mimic the electronic transport characteristics observed at the single molecule - level. The stability of these networks in air also offers a potential for building robust and efficient molecular junctions.

## Experimental

### Synthesis of organophilic AuNPs and functionalisation

Organophilic AuNPs were synthesised by adding 10 mg of gold chloride hydrate (HAuCl4).3H2O (99.9%, Sigma Aldrich) to a mixture (1:1 v/v) of oleylamine (technical grade, 70%, Merk) 1.5 ml and anhydrous toluene (99.8%, sigma Aldrich) 1.5 ml. The mixture was heated at 200 °C for 45 mins. After the completion of the reaction, which is evident by the formation of dark red colour, the mixture was added in methanol and the nanoparticles were separated by centrifugation and thoroughly washed with methanol or ethanol at least four times in order to remove any excess of surfactants (OAm). The Au NPs were subsequently solubilised in hexane (anhydrous, 95%, Sigma Aldrich). 100 µL of l-dodecanethiol (≥98%, Sigma Aldrich) was added to the hexane dispersion of oleylamine -capped AuNPs. The resulting solution was stored in the fridge for 3 days to precipitate. The supernatant was removed and the nanoparticles were redispersed in hexane

### Fabrication of the 2D network and molecular exchange

2D arrays of DDT-coated AuNPs are prepared at the air/water interface. The assembly is carried out in a Teflon beaker, which provides slightly convex water surface Typically, a few drops of the hexane (~ 300 µL) solution containing DDT-capped AuNPs are spread on water surface. using a pipette. During the evaporation of hexane, DDT capped AuNPs arrange themselves into a close-packed array. The array was then transferred to a Si/SiO2 substrate (consisting of 80 pairs of interdigitated (ID) electrons with 500 nm interelectrode spacing) by micro-contact printing using a poly-dimethylsiloxane (PDMS). Stamp was blow-dried with nitrogen gun prior to stamping to ensure all water droplets were removed. The investigated di(SAc) molecules (for synthesis and characterisation details of the di(SAc) molecules, see supplementary information) were introduced into the arrays by place exchange. To achieve this, samples were immersed in 2 µM solution of the incoming molecule in hexane over the course of 5-15 days. Samples were then taken out of solution and dried with N2 gas for electrical measurements.

**Characterisation**

The IV measurements were performed on a TTPX probe station in air at room temperature. Measurement of temperature dependence of electrical properties of the arrays were done using Cryogen free high field system between 300K and 3K.

UV-Vis absorption spectra were measured using Shimadzu UV-2600 spectrometer in the range 200-800 nm with 2 nm resolution. Dynamic light scattering (DLS) measurements of the nanoparticles in hexane were carried out using Malvern Panalytical ZetaSizer Ultra. Samples for Transmission Electron Microscopy (TEM) were prepared by placing a drop of AuNPs/DDT solution on a carbon coated copper grid. TEM was performed using a JEOL- JEM 1230 TEM. XPS measurements were conducted using a Thermo Fisher Scientific Nexsa XPS. All data were acquired using monochromatic Al Kα X-rays (1486.6 eV), with a pass energy of 25 eV. No considerable sample charging was observed. Charge referencing was done against adventitious carbon (C 1s, 284.6 eV). Spectra are presented with intensity in counts per second (CPS) without smoothing. Fitting of the experimental peaks were obtained using combinations of Gauss-Lorentz Mix Product lines. Applying Smart type background substruction using the Thermofisher Scoentofic Avantage software. UPS analyses were conducted on a Thermo Fisher Scientific Nexsa XPS. UPS spectra were obtained with a He lamp, He I of hv = 21.2 eV and I=60mA. To separate the secondary edges of the sample and analyser, a potential of 10 V was applied to the sample. The spectra were recorded in the range from -0.78 to 21.2eV, with energy step size of 0.05eV, pass energy of 2eV, and dwell time of 50ms.

## Result and Discussion

**Characterisation of Functionalised AuNPs and Assembly** As-prepared AuNPs exhibit surface plasmon resonance (SPR) peak at about 521 nm in the UV-Vis spectra (Figure 2). The SPR peak of AuNPs shifts to slightly longer wavelength ~ 530 nm after the addition of DDT. In contrast, AuNPs with the di(SAc)-terminated compounds exhibit blue shift with characteristic absorption peaks at

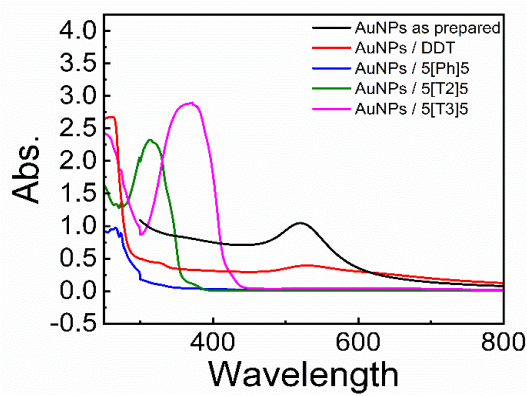

Figure 1. UV-Vis absorption spectra of hexane dispersions of as prepared AuNPs and AuNPs protected by DDT and di(SAc)-terminated molecules 5[Ph]5, 5[T2]5 and 5[T3]5. The solutions were prepared by adding 200 µL of a hexane suspension of 2.5 mM AuNPs to a solution of 2µ mole of the desired surfactant in 10 mL hexane.

368 nm, 315 nm, and 267 nm for 5[T3]5 and 5[T2]5 and 5[Ph]5, respectively. These are also consistent with the presence of α-ter- and α-bi thiophene,[20] and phenyl moieties within these molecules. The observed hyperchromic shift for the solutions containing 5[T3]5 and 5[T2]5 may be attributed to extended pi-conjugation. The absence of a SPR band in the UV-vis spectra for the di(SAc) compounds may indicate that the surface electrons could be localised, as may be expected from the strong Au-S covalent bonding.

The formation of ordered monolayers of AuNPs is confirmed by TEM. Figure 3a shows a TEM micrograph of a monolayer of spherical AuNPs coated with DDT. AuNPs form large areas of close packed hexagonal arrays where one particle is surrounded symmetrically by six other particles. The average size of gold nanoparticles in the 2D array is 6.8 nm (Figure 3c), and the average centre-to-centre distance between two adjacent nanoparticles is 8.7 nm, The average gap between two nanoparticles is 1.9 nm, which is very close to the length of the DDT tail ~1.7 nm. The narrow size distribution was also confirmed by DLS measurements (Figure 3b). The as-prepared AuNP shows only one peak with a narrow distribution, suggesting a monodispersed AuNP suspension with no agglomeration. The intensity-weighted average particle diameter (~11 nm) is slightly greater than the value obtained from TEM. The difference can be rationalised by DLS being a technique sensitive to coordinated ligands and surfactant and therefore providing a fundamentally different measure of particle size.

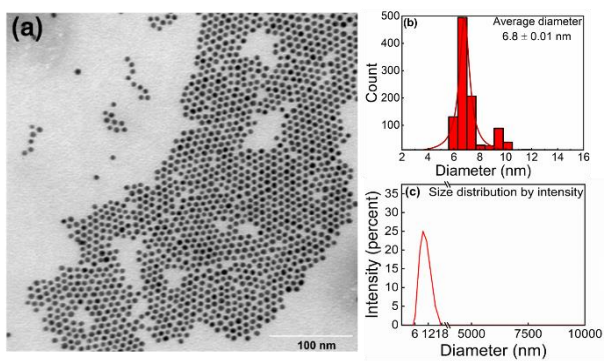

Figure 3. **(a)** TEM image of AuNPs/DDT forming a monolayer ordered array in hcp superlattice. **(b)** histogram of TEM size distribution. **(c)** analysis of size distribution by DLS showing the averaged (3 measurements per sample) intensity size distribution for as-prepared AuNPs

XPS was used to probe the surface chemistry of the functionalised AuNPs after molecular replacement with the di(SAc) compounds. Full survey spectra were recorded and the signals for C 1s and Au 4f are presented in Figure 4, meanwhile O 1s, and S 2p signals can be found in the supplementary information (Figure S1). The sets of binding energy positions for C 1s and Au 4f core levels are given in Table S1. In all samples, the C 1s XPS peaks were deconvoluted into 3 peaks using a Gaussian-Lorentzian peak profile. The binding energy of the C-C and C-H chemical bonds are assigned at 283.88 eV-285.67 eV. The chemical shifts to higher binding energy values 285.92 -287.19 eV are due to C-S or C-O/C=O species.[21-23] The observed peak positions situated near 283 eV represent the sp2 hybridized state of carbon atoms (the binding energy of π bond is normally ~ 1eV less than the binding energy of σ bond).[24] Two major peaks due to two spin-orbit components Au 4f7/2 and Au 4f5/2 at ~ 82.70 - 84.19 eV and 86.50 - 88.02 eV, respectively were observed. These values match well to reference binding energy of metallic AuNPs with zero oxidation state.[25-28] The curve-fitting of the Au 4f components manifests another contribution at higher binding energy values (~ +1 ev, Table S1) which can be associated with gold atoms that are covalently bound to sulfur.[23] The sulfur in all samples shows weak, broad signals, in the range 160 -170 eV (supplementary information Figure S1), signalling a very thin adsorbed layer of the organic molecule. We found S 2p doublets for all the SAMs at binding energies ~ 161 -163 eV, which are in agreement with the values that are usually observed for various sulfur species bound to the surface of gold.[29, 30] A second S 2p doublet was observed at the position 164-165 eV in the samples prepared using 5[T2]5 and 5[T3]5, representing the thiophene moiety within the structure of these molecules that is not bound to gold.[31] However, the 5[Ph]5 coated samples showed S 2p doublets at higher binding energies 167-168 eV, which may be attributed to oxidized sulfur, such as in sulfones or sulfoxides arising from possible reactions related to water contamination.

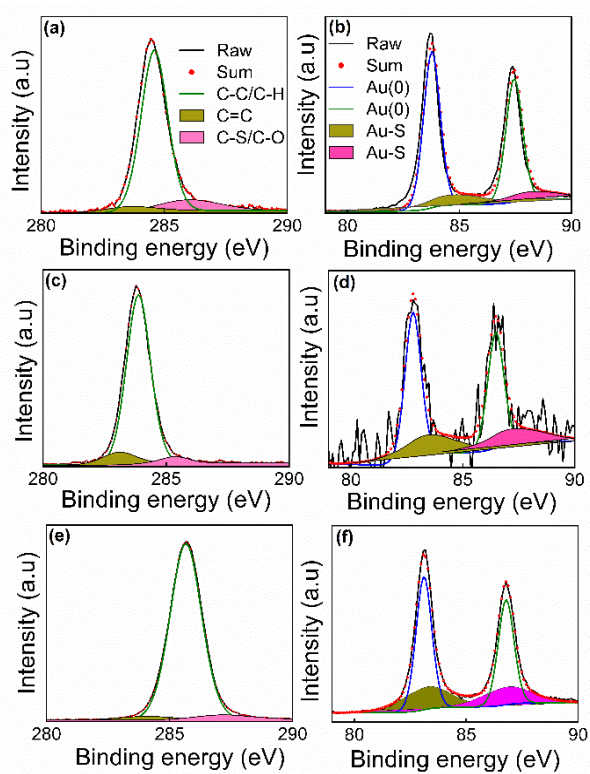

Figure 4. C 1s (**a**, **c**, **e**) and Au 4f (**b**, **d**, **f**) XPS signals for SAMs from AuNPs functionlised with 5[T3]5 (**a**, **b**), 5[T2]5 (**c**, **d**), and 5[Ph]5 (**e**, **f**).The C 1s and Au 4f $_{7/2}$ peaks were deconvoluted to estimate the sp$^3$ , sp$^2$ and Au-S bonding in the prepared SAMs.

**Electrical transport measurements**

One major motivation behind the current study is to provide insight into charge transport through SAM of AuNPs interlinked by di(SAc)-terminated molecules. In contrast to single molecule junction studies, statistics of experimental conductance of self-assembled monolayers of molecule-metal-molecule are rare in the literature. In this work, room-temperature conductance histograms were constructed from individual current-voltage (*IV*) traces. Control measurements on substrates without monolayers show no noticeable (≤ 10-13 A) current in the range ± 20 V. All devices exhibited non-linear IV characteristics over the ± 20 V bias range as shown in Figure 5a for a typical device. The exponential increase of current with increasing bias is consistent with a tunnelling mechanism of charge transport through the nanoparticle-molecule junctions.[32, 33] To test this assumption, the temperature dependence of conductance was evaluated for AuNPs coated with DDT and 5[Ph]5. Figures 5b and 5c reveal that at low temperatures, both arrays showed clear voltage threshold for conduction, signalling strong Coulomb blockade of transport.[34] Additional current paths are opened at voltage beyond the threshold, leading to linear IV behaviour in which sequential transport is the dominant mechanism. Figure 5a also shows that as the temperature rises, the threshold for lifting the blockade regime shifts to lower voltage.

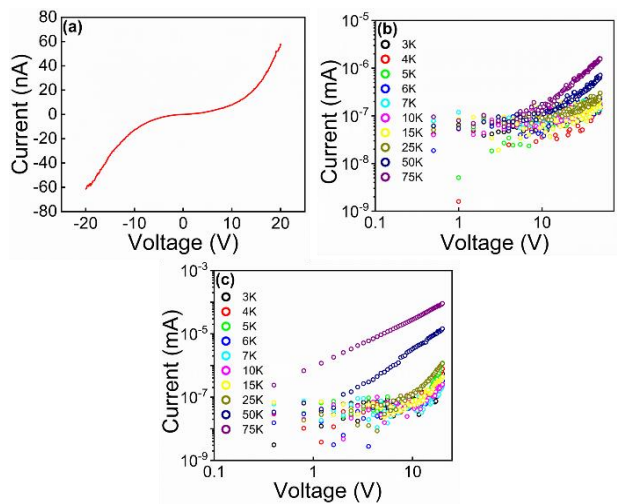

Figure 5. (a) IV relation of SAM AuNPs of a representative device with 500 nm electrodes separation. (b) and (c) low temperature dependence of IV plots for SAM AuNPs capped with DDT (b) and 5[Ph]5 (c) plotted on a double log scale. Due to low conductance of AuNPs/DDT device was measured up to 50V.

Room-temperature conductance histograms for the self-assembled AuNP pre-exchange (coated with DDT) and post exchange for 5[Ph]5 (37 devices), 5[T3]5 (63 devices) and 5[T2]5 (63 devices) are presented in Figure 6. The conductance values of all devices were calculated by determining the slope of the linear region of the IV curve. The investigated molecular wires are different on the basis of their extent of conjugation electron density, which in turn affects molecular conductance. Table 3 summarises the average conductance of the investigated molecules before and after exchange obtained from fitting the histograms to Gaussian functions. AuNP/DDT has an average conductance value in order of pS. The slight sample to-sample variation in AuNP/DDT monolayers may be due to changes in the network topology. The average conductance value increased one order of magnitude and reached saturation after 5 days' replacement with molecule 5[Ph]5 ~ 2.96 X $10^{-11}$ S. Meanwhile, we observed a two orders of magnitude increase for the larger molecules 5[T2]5 and 5[T3]5 after 15 days' replacement. The highest average conductance is observed for AuNPs cross-linked with 5[T2]5.

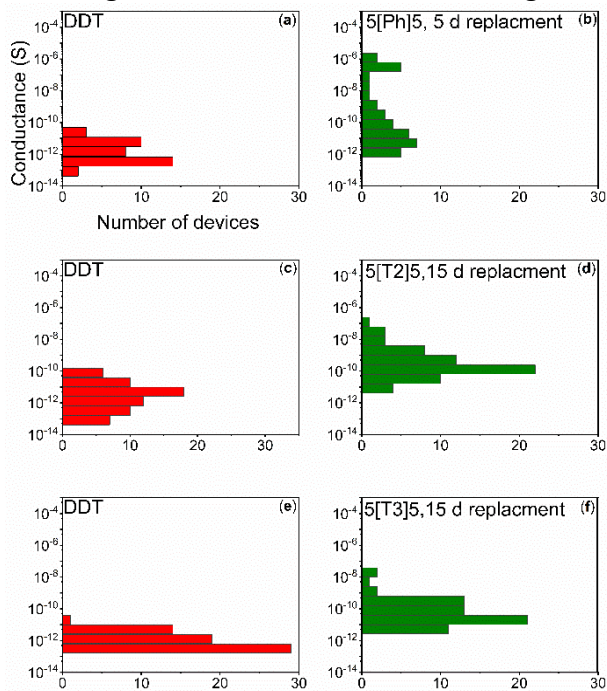

Figure 6. Conductance of SAM AuNPs versus occurrence in a log-normal histogram pre-exchange (**a**, **c**, **e**) and post-exchange for 5[Ph]5 (**b**,), 5[T2]5 (**d**) and 5[T3]5 (**f**).

Table 1. Summary of the average conductance values for SAM AuNP monolayers pre-exchange and post-exchange (17 hrs, 5 days and 15 days' exchange). Values are in units of Siemens (S). Conductance was obtained from Gaussian fits to conductance histograms. The conductance histograms for 17 hrs and 5 days' replacement are provided in supplementary information Figure S2.

| Sample | Pre - exchange | 17 hrs exchange | 5 d exchange | 15 d exchange |
|---|---|---|---|---|
| 5[Ph]5 | $1.89 \times 10^{-12}$ | $2.83 \times 10^{-12}$ | $2.96 \times 10^{-11}$ | - |
| 5[T2]5 | $7.06 \times 10^{-12}$ | - | - | $7.09 \times 10^{-10}$ |
| 5[T3]5 | $4.87 \times 10^{-12}$ | - | $1.25 \times 10^{-11}$ | $1.003 \times 10^{-10}$ |

**Electronic structure and Theoretical Conductance**

Experiment and theory were combined to understand the conductance behaviour of the di(SAc) SAMs. The surface electronic structure of the materials was probed using UPS which allows direct measure of materials work functions and valence band maxima (HOMO). UPS spectra of all SAMs are shown in Figure 7a. The work function is extracted from the difference between the incident UV photon energy (21.21 eV) and the binding energy of the secondary electron edge (SEE) at the high binding energy region of the UPS spectra (determined by a linear extrapolation of the secondary electron cutoff), taking into account that the Fermi energy is set to zero binding energy. As illustrated in Figure 7b, in all cases, the work functions of the SAMs is decreased from that of bulk Au (5.2 eV). The replacement of DDT (work function = 5.04 eV) with 5[Ph]5, 5[T2]5 and 5[T3]5 decreases the work function to 4.43 eV, 4.93 eV and 4.86 eV respectively. This means that the electron injection barrier in to the LUMO level is reduced for the di(SAc) compounds. The optical band gap is evaluated using the equation $E_g = 1240/ \lambda_{edg}(nm)$ where $\lambda_{edg}$ is the onset value of the absorption spectrum (UV-Vis) at the long wavelength region. The LUMO position is then determined from the HOMO and band gap value. Table 2 provides a summary of the band structure and energy levels for the investigated SAMs. The location of the energy levels of the molecule relative to the Fermi energy of the metallic contact is an important factor in determining the conductance of molecular arrays. Therefore, based on the obtained values, we reconstructed a band structure for the compounds used in this study, as shown in Figure 7c. The band model scheme demonstrates that gold Fermi level lies in between HOMO-LUMO gap for all molecules. It also predicts that the HOMO level of the 5[T2]5 compound is closer to Au Fermi energy, which may explain its enhanced conductance.

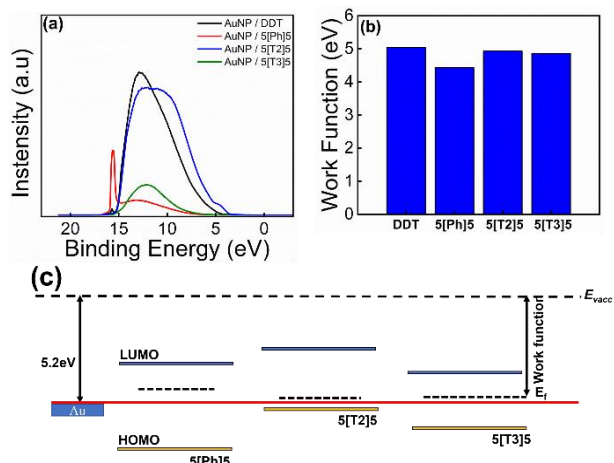

Figure 7. **(a)** UPS spectra of DDT and di(SAc) SAMs. **(b)** the extracted work functions. **(c)** schematic of the electronic structure of the investigated SAMs.

Table 2. A summary of the band structure and energy levels for SAMs of AuNPs functionalised with 5[Ph]5, 5[T2]5 and 5[T3]5. The HOMO edge position is obtained by a linear extrapolation at the low binding energy cutoff region of the UPS spectra. Units are eV

| Sample | Eg | $E_f$ | HOMO | LUMO |
|---|---|---|---|---|
| 5[Ph]5 | 4.20 | 4.43 | 7.21 | 3.01 |
| 5[T2]5 | 3.46 | 4.93 | 5.35 | 1.89 |
| 5[T3]5 | 2.94 | 4.86 | 6.26 | 3.32 |

Density functional theory (DFT) was used to compute electrical conductance and explore the interplay between electrical conductance of different molecular wire and their central moiety. We first perform the geometry optimisation of molecules between gold electrodes using the SIESTA[35] implementation of DFT and obtain the ground-state mean-field Hamiltonian and overlap matrix elements. These results were then combined with the Gollum.[36, 37] Implementation of the nonequilibrium Green's function method to compute the electrical conductance of the junctions (see further details in supplementary information).The presence of a central unit in the alkyl chain induces conformational changes, leading to various configurations and binding conformation to the electrodes and therefore modifying the electronic structure of the molecule.[18] For this reason, we calculate the electrical conductance of five different energetically optimised binding configurations of each molecule as shown in Figure 8.

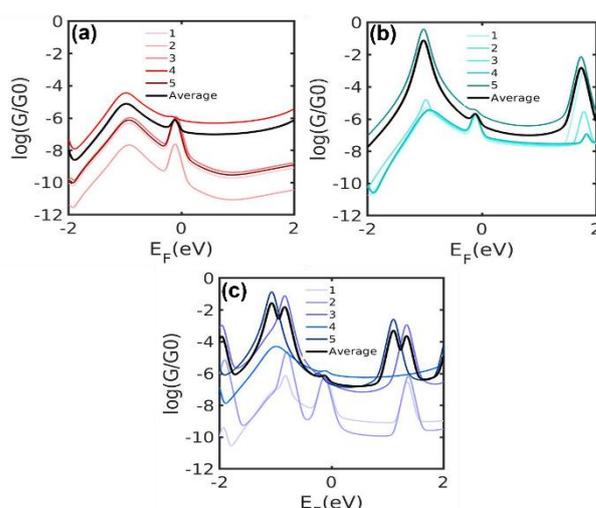

Figure 8. Calculated conductance of the molecule: **(a)** 5[Ph]5, **(b)** 5[T2]5 and **(c)** 5[T3]5 connected to electrodes using five different configurations.

The relaxed molecular structure of single molecular junction for 5[Ph]5, 5[T2]5 and 5[T3]5 respectively and the average of electrical conductance values of five different configurations for each molecule are shown in Figure S3. Typically, the Fermi energy, EF, lies near the middle of the HOMO–LUMO gap, and comparing with the experimental trend our results suggest that EF falls within the highlighted region in the figure. The calculation with different binding configurations to electrodes for the di(SAc) molecules indicates that electron transport through the junctions varies according to the different configurations. The overall trend for the average conductance of different configurations) agrees well with our experiment for a wide range of Fermi energies around DFT Fermi energy (5[T2]5 > 5[T3]5 > 5[Ph]5). The frontier molecular orbitals and energies are listed in Table S2. The predicted band gaps are in close agreement to our experimental values, however, The HOMO energy level of 5[T2]5 (-5.61 eV) is slightly higher than that of 5[T3]5 molecule (-5.36 eV). Generally, the frontier orbitals relative to the Fermi energy of Au contacts is not necessarily accurately determined by DFT calculations, which therefore does not allow a reliable comparison with the experimental HOMO energy levels obtained from UPS. Furthermore, discrepancy between experiment and theory might originate from the interactions with other molecules and nanoparticles in the assembly which is lacking in the single molecule calculations. Clearly, Different molecule-electrode contact geometries lead to large variations in electrical conductance. This is because the electronic coupling between Au-S and central group is significantly affected by confirmation of molecule.

## Conclusions

We have successfully incorporated di(SAc)-terminated molecules into self-assembled arrays of DDT-functionalised AuNPs. Cross-linking and conjugation are proposed as a viable way to improve charge-transport efficiency The final conductance is enhanced by up to two orders of magnitude for the di(SAc)-terminated molecules. The results demonstrate that the single-molecule central unit effects can be translated into ordered 2D assemblies of AuNPs. The presence of conjugated moiety in the alkyl chain affects the final conductance of the array, the magnitude of which depends on the nature of the central unit. This work extends the field by exploring a new series of dithioacetate di(SAc) molecules and provides insight into their electronic behaviour.

## Author Contributions

M. A and J. A. M. conceived and designed the experiments. A.V. synthesised the organic molecules and characterised them. M. A. synthesised the samples, fabricated devices and performed structural and electrical characterisation with the help of J. B. and E. P. A. D., S. S. and H. S. developed the theoretical explanation, and A. D. performed the calculations. M.A. analysed the results with the help of A. V., S. J. H and J. A. M. M. A wrote the manuscript with the help of all co-authors

## Conflicts of interest

The authors declare no competing interests.

## Acknowledgements

J. A M. was supported through the UKRI Future Leader Fellowship, Grant No. MR/S032541/1, with in-kind support from the Royal Academy of Engineering. M.A. acknowledges the NanoVision Centre at QMUL for using the TEM facility.

# References

1.	C. B. Murray, C. R. Kagan and M. G. Bawendi, Annual Review of Materials Science, 2000, **30**, 545-610.

2.	D. V. Talapin, J. S. Lee, M. V. Kovalenko and E. V. Shevchenko, Chemical Reviews, 2010, **110**, 389-458.

3.	B. Pelaz, S. Jaber, D. J. de Aberasturi, V. Wulf, T. Aida, J. M. de la Fuente, J. Feldmann, H. E. Gaub, L. Josephson, C. R. Kagan, N. A. Kotov, L. M. Liz-Marzan, H. Mattoussi, P. Mulvaney, C. B. Murray, A. L. Rogach, P. S. Weiss, I. Willner and W. J. Parak, Acs Nano, 2012, **6**, 8468-8483.

4.	Z. H. Nie, A. Petukhova and E. Kumacheva, Nature Nanotechnology, 2010, **5**, 15-25.

5.	M. C. Daniel and D. Astruc, Chemical Reviews, 2004, **104**, 293-346.

6.	S. Q. Liu and Z. Y. Tang, Journal of Materials Chemistry, 2010, **20**, 24-35.

7.	J. H. Liao, J. S. Agustsson, S. M. Wu, C. Schonenberger, M. Calame, Y. Leroux, M. Mayor, O. Jeannin, Y. F. Ran, S. X. Liu and S. Decurtins, Nano Letters, 2010, **10**, 759-764.

8.	J. H. Liao, S. Blok, S. J. van der Molen, S. Diefenbach, A. W. Holleitner, C. Schonenberger, A. Vladyka and M. Calame, Chemical Society Reviews, 2015, **44**, 999-1014.

9.	Q. Chen, X. D. Liu, J. F. Zeng, Z. P. Cheng and Z. Liu, Biomaterials, 2016, **98**, 23-30.

10.	W. B. Chang, B. Russ, V. Ho, J. J. Urban and R. A. Segalman, Physical Chemistry Chemical Physics, 2015, **17**, 6207-6211.

11.	H. Wang, L. Yao, X. Mao, K. Wang, L. Zhu and J. Zhu, Nanoscale, 2019, **11**, 13917-13923.

12.	T. Taguchi, K. Isozaki and K. Miki, Advanced Materials, 2012, **24**, 6462-6467.

13.	L. Srisombat, A. C. Jamison and T. R. Lee, Colloids and Surfaces a-Physicochemical and Engineering Aspects, 2011, **390**, 1-19.

14.	E. Leary, S. J. Higgins, H. van Zalinge, W. Haiss, R. J. Nichols, S. Nygaard, J. O. Jeppesen and J. Ulstrup, Journal of the American Chemical Society, 2008, **130**, 12204–12205.

15.	Xiao, Xu and N. J. Tao, Nano Letters, 2004, **4**, 267-271.

16.	M. P. Samanta, W. Tian, S. Datta, J. I. Henderson and C. P. Kubiak, Physical Review B, 1996, **53**, R7626-R7629.

17.	J. Liao, M. A. Mangold, S. Grunder, M. Mayor, C. Schönenberger and M. Calame, Journal, 2008, **10**, 065019.

18.	S. Sangtarash, A. Vezzoli, H. Sadeghi, N. Ferri, H. M. O'Brien, I. Grace, L. Bouffier, S. J. Higgins, R. J. Nichols and C. J. Lambert, Nanoscale, 2018, **10**, 3060-3067.

19.	L. Bernard, Y. Kamdzhilov, M. Calame, S. J. van der Molen, J. H. Liao and C. Schonenberger, Journal of Physical Chemistry C, 2007, **111**, 18445-18450.

20.	K. Wang, A. Vezzoli, I. M. Grace, M. McLaughlin, R. J. Nichols, B. Q. Xu, C. J. Lambert and S. J. Higgins, Chemical Science, 2019, **10**, 2396-2403.

21.     Y. Joseph, I. Besnard, M. Rosenberger, B. Guse, H. G. Nothofer, J. M. Wessels, U. Wild, A. Knop-Gericke, D. S. Su, R. Schlogl, A. Yasuda and T. Vossmeyer, Journal of Physical Chemistry B, 2003, **107**, 7406-7413.

22.     D. J. Morgan, C-Journal of Carbon Research, 2021, **7**, 51.

23.     F. Vitale, I. Fratoddi, C. Battocchio, E. Piscopiello, L. Tapfer, M. V. Russo, G. Polzonetti and C. Giannini, Nanoscale Research Letters, 2011, **6**, 9.

24.     T. R. Gengenbach, G. H. Major, M. R. Linford and C. D. Easton, Journal of Vacuum Science & Technology A, 2021, **39**, 013204.

25.     M. C. Bourg, A. Badia and R. B. Lennox, Journal of Physical Chemistry B, 2000, **104**, 6562-6567.

26.     L. Mohrhusen and M. Osmic, Rsc Advances, 2017, **7**, 12897-12907.

27.     Z. Y. Huo, C. K. Tsung, W. Y. Huang, X. F. Zhang and P. D. Yang, Nano Letters, 2008, **8**, 2041-2044.

28.     R. Radnik, C. Mohr and P. Claus, Physical Chemistry Chemical Physics, 2003, **5**, 172-177.

29.     D. G. Castner, K. Hinds and D. W. Grainger, Langmuir, 1996, **12**, 5083-5086.

30.     S. Zhang, G. Leem and T. R. Lee, Langmuir, 2009, **25**, 13855-13860.

31.     T. M. Jiang, W. Malone, Y. F. Tong, D. Dragoe, A. Bendounan, A. Kara and V. A. Esaulov, Journal of Physical Chemistry C, 2017, **121**, 27923-27935.

32.     R. Parthasarathy, X. M. Lin and H. M. Jaeger, Physical Review Letters, 2001, **87**, 186807.

33.     K. Elteto, X.-M. Lin and H. M. Jaeger, Physical Review B, 2005, **71**, 205412.

34.     T. B. Tran, I. S. Beloborodov, J. Hu, X. M. Lin, T. F. Rosenbaum and H. M. Jaeger, Physical Review B, 2008, **78**, 075437.

35.     J. M. Soler, E. Artacho, J. D. Gale, A. García, J. Junquera, P. Ordejón and D. Sánchez-Portal, Journal of Physics: Condensed Matter, 2002, **14**, 2745-2779.

36.     H. Sadeghi, Nanotechnology, 2018, **29**, 373001.

37.     J. Ferrer, C. Lambert, V. Garcia-Suarez, D. Manrique, D. Visontai, L. Oroszlany, R. Rodriguez-Ferradas, I. Grace, S. Bailey, K. Gillemot, H. Sadeghi and L. Algharagholy, New Journal of Physics, 2014, **16**, 093029.